# A new approach to ultrasensitive gravitational wave detection

A. Gulian[a], J. Foreman[b], V. Nikoghosyan[a,c], L. Sica[a], J. Tollaksen[a], S. Nussinov[a,d]

[a]*Chapman University, Institute for Quantum Studies, Orange CA, 92866 & Burtonsville MD, 20866, USA*
[b]*Independent Researcher, Alexandria VA, 22310, USA*
[c]*Physics Research Institute, National Academy of Sciences, Ashtarak, 0203, Armenia*
[d]*School of Physics and Astronomy, Tel-Aviv University, Ramat-Aviv, 69978, Tel-Aviv, Israel*

**Abstract**

We suggest here a method to detect gravitational waves (GW) different from the interferometric approach. It is based on two critical steps: conversion of the GW action into rotational motion and subsequent conversion into electric current. The ability to detect extremely tiny currents empowers this suggested approach in which the gravitational wave moves charges around closed loops. This new method may allow detection of gravitational waves with strain sensitivities beyond the reach of the interferometric approaches.

### I. The merits of GW detectors based on induced currents relative to interferometry

Our novel concept which has matured since 2011 (consecutive steps can be traced as v1-v4 versions of Ref. [1]) is represented below in fairly simple and straightforward terms which should allow easy theoretical validation.

The driving force behind our interest is the great scientific merit of detecting gravitational waves from many diverse sources. The epochal recent discovery of LIGO [2], the first direct verification of the existence of gravitational waves by terrestrial, man-made means, is the successful culmination of efforts to achieve this Holy Grail over a very long time [3]. Along with earlier indirect results [4], it is a big step towards using GWs to investigate fundamental physics: GWs freely propagate and therefore constitute a unique new window allowing researchers to probe the Universe at its earliest hottest and densest stages [5].

The more direct motivation of our effort is the realization that the large (Avogadro-like) number of electrons in solids or Cooper pairs in superconductors may strongly amplify the effect of the extremely tiny physical motion induced by the GW. To our knowledge, no previous experimental designs using *both* of the above two steps have been described. While there have been several suggestions to use superconducting elements as basic parts of GW detectors (see, *e.g.,* [6]), they did not come to fruition.

A basic feature of GWs is that they change the relative separation between two probe masses proportional to the distance $L$ between them. This change, $\Delta L = L \cdot h_{GW}$, is incredibly small due to the small value of the GW amplitude $h_{GW}$. Specifically, the $h_{GW}$ in the recent LIGO observation was about $10^{-21}$ which means that (for $L = 4km$) $\Delta L$ at the level of $10^{-12}$ *micrometers* has been detected. Further increase of the sensitivity and frequency range of GW detectors will allow finding not only many more mergers of massive black holes of the type seen, but also GWs from many other more conventional or putative new physics [7] sources.

Can even weaker $h_{GW}$ be detected in other ways, especially using smaller size detectors? To answer this question all possible approaches should be considered and the many related physics issues should be analyzed.

Suppose that $\Delta L$ above is associated with a characteristic frequency $\omega$. Then the smallness of the velocity of the oscillatory motion $v = \omega \Delta L$ reflects the smallness of $\Delta L$: e.g., $v = 10^{-15} m/sec$ for $\omega = 100\ sec^{-1}$. However, if the ensemble of conducting electrons in metals is moving with this velocity, this results in a rather large current. Indeed, substitution of a typical value for the charge carriers in metals $n = 10^{29} m^{-3}$ yields a current $J = evnS \sim 0.1\ microamp$ for a conductor with cross section $S = 10 \times 10 cm^2$. Note that a current of $1 \mu A$ corresponds to $10^{13}$ electrons passing through the conductor's cross section per second. Single-electron accuracy in the detection of current is presently achievable (recall the existence of single-electron transistors). Thus, in the example above, one has ten orders of magnitude to spare. This would allow a dramatic decrease in detector size relative to LIGO and at the same time a strong increase in sensitivity. The large number of electrons (reflecting the large Avogadro number) thus plays an essential role in the suggested approach.

Inducing currents by GWs is naturally achieved by translating their effect into a mechanical rotational motion.

## II. Obtaining rotational motion from GW

Consider a massive circular ring in the x,y-plane perpendicular to the z-direction of the incident GW. For a pure-polarized GW[1] the system of coordinates can always be oriented as shown in Fig. 1 [8]. In local nearly flat space-time, the GW force acting on a particle with mass $m$ at location $r = (x_0, y_0)$ is [9]:

$$F^i_{GW}(r,t)/m \equiv a^i_{GW}(r,t) = -R^i{}_{0j0} r^j / 2, \qquad (1)$$

where $i, j = (x, y)$, and the relevant components of the Riemann tensor are given in terms of the transverse traceless perturbation of the metric:

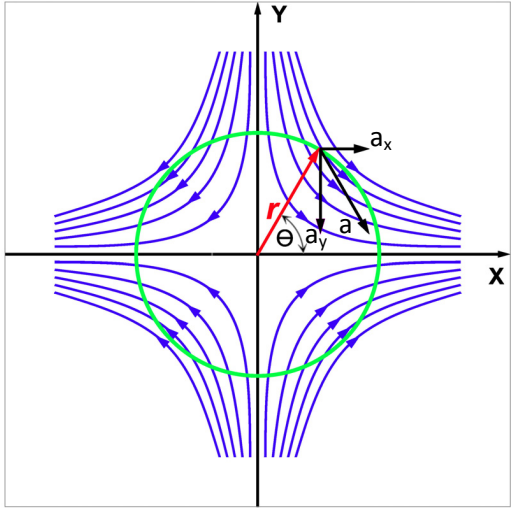

FIG. 1. The acceleration **a** of test particles on the ring. GW field-lines are indicated by the set of hyperbolas. As shown in the text, everywhere on the circle, |**a**| is constant. The direction of **a** at a given point **r** of the circle is tangential to the GW field-line crossing or touching the circle at that point.

$$R^x{}_{0x0} = -h^{TT}_{xx,00}/2 = -R^y{}_{0y0} = \omega^2 h^{TT}_{xx}/2, \qquad (2)$$

---

[1] In case of a general polarization of the GW, the response is a simple superposition.

$$R^y{}_{0x0} = R^x{}_{0y0} = 0. \tag{3}$$

Equations (2) and (3) are valid for the chosen coordinate system and for a harmonic GW with frequency $\omega$: $h^{TT}_{xx} = h_{GW}\cos(\omega t)$, so that $a^x_{GW}(\mathbf{r},t) = -(\omega^2 x_0 h_{GW}/4)\cos(\omega t)$, and $a^y_{GW}(\mathbf{r},t) = (\omega^2 y_0 h_{GW}/4)\cos(\omega t)$. When the masses constitute a solid the deformation of the ring can usually be neglected (the utilization of the deformation of conductors for GW detection in exotic cases is considered in [10,11]). The contribution to the torque of each mass element on the circle $dm(\mathbf{r}) = dm(\theta) = \rho(\theta)d\theta$, where $\rho(\theta)$ is the linear density, is
$\tau_z(\mathbf{r}) = \tau_z(\theta) = x_0 F_y - y_0 F_x = -rF\sin(2\theta) = -dm(\theta)\omega^2 h_{GW} r^2 \sin(2\theta)\cos(\omega t)/4$. The total torque is then an integral over the angle $\theta$. In the simplest case of homogeneous mass distribution the torque vanishes by symmetry. However, lower symmetry mass distributions, like $\rho(\theta) = \rho_A$ for $\{0 < \theta < \pi/2;\ \pi < \theta < 3\pi/2\}$ and $\rho(\theta) = \rho_B$ for $\{\pi/2 < \theta < \pi;\ 3\pi/2 < \theta < 2\pi\}$, with $\rho_A \neq \rho_B$ as in Fig. 2, result in non-zero torque relative to unchanged center of mass. The total torque imposed by GW is then $\tau_z^{GW} = (M_B - M_A)\omega^2 h_{GW}\cos(\omega t)r^2/4$, where $M_A$ and $M_B$ are total masses of the $A$ and $B$ quadrants. Since the moment of inertia in this case is $I = (M_A + M_B)r^2$, the angular velocity $\Omega$ can be found from the equation of motion

$$\frac{d\Omega}{dt} = \frac{\tau_z^{GW}}{I} = \frac{M_B - M_A}{4(M_B + M_A)}\omega^2 h_{GW}\cos(\omega t) \approx \frac{1}{4}\omega^2 h_{GW}\cos(\omega t), \tag{4}$$

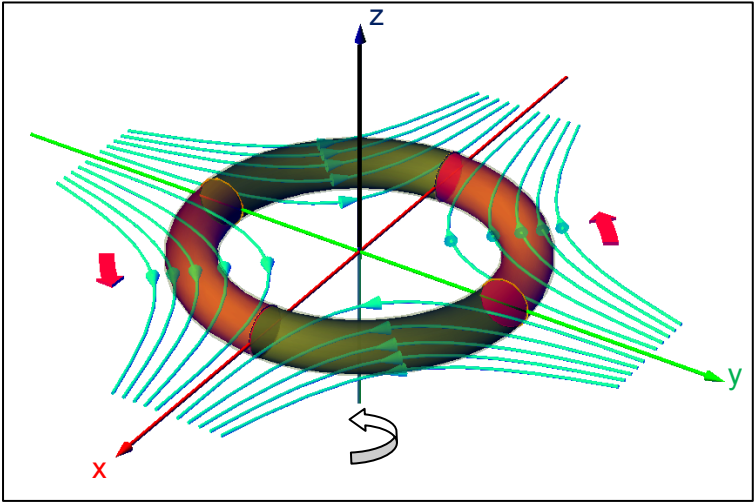

FIG. 2. The gravitational field lines are the same as in Fig. 1. If the mass distribution in the torus is uniform then GW will only distort it. No net rotation will result as the torques generated in the four quadrants identically cancel. However, since we make the two diametrically opposite quadrants, which are colored in red, much heavier than the other pair, the dominant torques in this case, indicated by the red arrows, will add up and generate a net rotation.

The direction of these arrows will reverse every half cycle causing torque reversal and thus oscillatory rotation of the torus around the z-axis.

where $M_A << M_B \equiv M$ is assumed, as

$$\Omega(t) = \frac{1}{4}\omega h_{GW}\sin(\omega t). \tag{5}$$

Hitherto it was tacitly assumed that the presence of finite distributed masses does not affect the GW field presented via Eqs. (2) and (3). This is expected from General Relativity, since the GW-field can always be expressed via the Weyl tensor, thus constituting that part of Riemann's tensor which does not depend locally on the mass distribution [12].

Therefore a circular oscillatory motion can be expected in the x-y plane, despite the GW field has no z-component of its (pseudo) curl[2]. This expectation is one of the two cornerstones of our approach. It may seem somewhat surprising and unexpected at first glance, but is clearly based on simple and rather elementary arguments. A similar conclusion was independently obtained in Ref. [14] (TOBA project in Japan) in which interferometric methods were suggested to detect the relative counter-rotation of two bars. In our case, the rotational oscillatory motion generated is converted into electric currents by a mechanism which will be described now.

### III. The simplest electric signal generation and resulting strain sensitivity

Many different schemes can be considered [15] for converting the rotational motion into electric signals. This conversion, underlying the generation of electricity, has been extensively discussed since Faraday's time and can be achieved with almost 100% efficiency. To present our idea in the simplest setting, consider a frame with a coil placed in a magnetic field as shown in Fig. 3. This frame is rigidly attached to the torus, and rotates around the z-axis together with the torus. Note that this figure represents the proposed apparatus as seen by the incident GW.

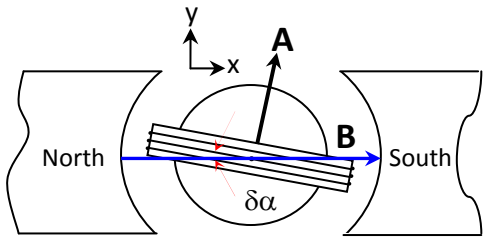

FIG. 3. The conversion of rotational motion into electric current. The homogeneous magnetic field **B** between the North and the South poles of the magnet is directed along the blue arrow. $\delta\alpha$ is the angle between the rotating frame and the **B**-field. The circle schematically indicates the quadrupolar torus described in Fig. 2.

In this "galvanometric" setup the magnetic flux through a single winding loop is
$\Phi(t) = \mathbf{B} \cdot \mathbf{A} = |\mathbf{B}| \cdot |\mathbf{A}| \cos\alpha(t)$, where **B** is the magnetic induction, and **A** is the vector perpendicular to the frame of magnitude $|\mathbf{A}| (\equiv A)$ equal to its area. Upon rotation of the frame the magnetic flux through the single loop changes as

$$\frac{d\Phi}{dt} = -B\,A \sin\alpha(t) \frac{d\alpha}{dt}. \tag{6}$$

For maximal sensitivity, **B** and **A** should be orthogonal ($\sin\alpha \approx 1$). Here $d\alpha/dt$ is caused by the rotation of the torus under the influence of GW (since the frame is rigidly attached to the torus) so as

$$d\alpha/dt \equiv \Omega(t) \cong (1/4)\,\omega\,h_{GW} \sin(\omega t). \tag{7}$$

In writing this relation we neglected the electromagnetic feedback, which will modify the solution (5). Later on (in Section VI) we will return to this issue and justify this approximation.

Equations (6) and (7) yield: $d\Phi/dt \cong (1/4) B\,A\,\omega\,h_{GW}\sin(\omega t)$. For the case where the coil consists of $N_0$ loops the corresponding amplitude of the flux oscillations is:

---

[2] One can recall a familiar analog in optics: filtering linearly polarized light with appropriate crystals can convert it into circular polarization – which carries net angular momentum along the direction of propagation – leaving atoms with the opposite angular momentum (see, *e.g.*, [13]).

$$\Phi_0 \cong (1/4) B N_0 A h_{GW}. \tag{8}$$

This flux is transferred in the usual manner via flux transformer circuit to a region away from the strong magnetic field where it will be amplified and measured by sensitive cryogenic electronics. In order to be measureable, the magnitude of the flux change should be comparable with that of the intrinsic noise of the suggested electronics. Even if the latter consists only of a single off-the-shelf SQUID, it will have a remarkably low noise level of $10^{-6} \phi_0 Hz^{-1/2}$ at 4K temperature [16], where $\phi_0 \cong 2 \times 10^{-15} Weber$ is the flux quantum. Demanding that the signal (8) be greater than this noise yields the minimum value of $h_{GW}^{min}$ detectable per unit bandwidth:

$$h_{GW}^{min} \sim 4 \times 10^{-6} \phi_0 / (N_0 BA). \tag{9}$$

For $B \sim 1\, Tesla$, $N_0 = 10^4$, and $A = a^2 \sim 1\, meter^2$, Eq. (9) yields a remarkable strain sensitivity of $10^{-24}\, Hz^{-1/2}$. More complex pick-up electronics with a large number of SQUIDs further enhances the signal-to-noise ratio. Independently, a reduction of the temperature from the above value 4K to 300mK is achievable, and it will further reduce the electronic noise.

In the above example a 1 $Tesla$ B-field was used. Such large fields can potentially generate large amounts of noise, which requires special precaution. However, as shown later (see Fig.4) much smaller fields may be used in detectors optimized for detecting GWs of specific frequencies.

## IV. Strain sensitivity and electromagnetic noise

In the above discussion it was assumed that no electromagnetic noise is picked-up. However, no perfect screening of electromagnetic noise is feasible. Screening is especially important when signals of the order of $\sim 10^{-6} \phi_0 Hz^{-1/2}$ are to be measured. Even a tiny amount of external electromagnetic noise penetrating the shielding could prevent detection of such small signals. However, this can be addressed by using the following general strategy. An auxiliary pick-up coil with almost identical size, shape, and makeup as the detecting coil, and therefore having a very similar response to electromagnetic radiation, is placed nearby with its plane parallel to that of the moving coil. It is affected by the same (electro) magnetic fields, but not attached to the moving quadrupolar frame, and therefore is not affected by the GW. By subtracting the output of this auxiliary coil from the signal in the detecting coil, one can strongly reduce the remaining spurious noise.

The described technique can complement the shielding, but intrinsic noise such as Johnson-Nyquist noise [17] still remains. The noise in the superconducting loop in Fig. 3 arises in this case from the non-superconducting (normal) electrons. The average noise current in this loop is[3]:

$$\langle J_{noise} \rangle = [4(k_B T / R_n) \delta\!f]^{1/2}, \tag{10}$$

where $k_B$ is the Boltzmann constant, and

$$R_n = (4 a \rho_n / S) \exp[\Delta / (k_B T)] \tag{11}$$

---

[3] Note that the square root of the resistance appears in the denominator of Eq.(10) rather than in the numerator. This is because we are interested in the resulting noise current $\langle J_{noise} \rangle = \langle U_{Johnson-Nyquist} \rangle / R_n$ rather than in the Johnson-Nyquist potential, which is proportional to the square root of the resistance $R_n$.

is the resistance of the normal component of the superconductor. In Eq. (11) $\rho_n$ is the resistivity of unpaired electrons in superconducting wire, $S$ is the wire cross section, and $\Delta = \Delta(T)$ is the BCS gap in the electronic spectrum [17,18]. Even though the Cooper-pairs in the superconductor move without resistance and shunt the above normal resistance, this normal resistance still contributes to the noise yielding equation (10) above.

This current (10) creates a fluctuating flux in the loop:
$\langle \Phi_{noise}^{loop} \rangle = L \langle J_{noise} \rangle \approx \mu_0 [(k_B T Sa/\rho_n) \delta f]^{1/2} \exp[-\Delta/(2k_B T)]$. This fluctuating flux is proportional to $S^{1/2}$. To minimize it, we make the natural choice $S \sim \lambda_L^2 \sim 10^{-10} cm^2$, where $\lambda_L \sim 10^{-5} cm$ is the London penetration depth. Choosing also $\rho \sim 10^{-7} \Omega \cdot m$ and substituting $T \sim 1K$ in the pre-factor yields the following equation:

$$\langle \Phi_{noise}^{loop} \rangle \approx 10^{-6} \phi_0 (\delta f)^{1/2} \exp[-\Delta/(2k_B T)] \quad . \tag{12}$$

Up to the exponential factor, this coincides with the SQUID noise above, Eq. (7). The exponential factor is always smaller than 1, so the noise in the loop is smaller than the SQUID noise (7). It is possible to make this noise (12) much smaller than the SQUID noise using a superconducting material with critical temperature much higher than the operational temperature, so that $\Delta \gg k_B T$.

### V. A replica method approach to counter spurious noise

Seismic noise is an issue for terrestrial devices. For interferometric detectors seismic noise attenuators have been developed with an attenuation factor $\chi < 10^{-10}$ [19]. The idea of using auxiliary devices described above for electromagnetic noise reduction can also complement the LIGO-type attenuator platforms to further attenuate seismic noise. Specifically, a second apparatus mechanically mimicking the GW sensitive device could be placed on the same platform. However, the torus of this mimicking device is uniform and therefore not sensitive to GW. The parameters of this auxiliary apparatus should be calculated and tested on a vibrating test-bed so as to ensure identical responses of the two apparatuses to the relevant seismic noise. Again, subtracting the signals of the two frames will reduce the remaining vibrational noise.

The above consideration neglects the noise introduced by the pivot holding the rotating part of the detector. Moreover, it neglects the restoring force which opposes the GW action. In principle, for a GW source of known frequency a high quality-factor pivoting can be used for resonant enhancement of the amplitude of oscillations [20]. However, this issue requires special analysis. Frictionless pivoting can be achieved in orbital implementations, or, as shown in Ref. [21], in superfluid helium bath. Frictionless pivoting will be assumed for simplicity in subsequent formulas in this article.

The same principle can be readily applied for the reduction of the pick-up electromagnetic noise and the noise caused by fluctuations of the applied large B-field. This requires only a replica of the superconducting signal coil which will not be attached to the mobile torus and hence its output is only due to the background electromagnetic fluctuations. By subtracting the output of the replica coil from that of the genuine signal coil the electromagnetic noise will be largely reduced.

### VI. Magnetic feedback effects and frequency response range

Even in absence of pivoting, there is a magnetic feedback effect which reduces the total torque. Neglecting this feedback can only be justified when the magnetic field $B$ is small. However, the strain sensitivity (9) is proportional to $B^{-1}$, so higher fields are preferable, and thus the feedback should be taken into account.

During the rotation, the up or down motions of sections of the conducting loop with velocity $v_y = \Omega a = \omega h_{GW} a$ of the frame relative to the **B**-lines generate a Lorentz force on the electrons $F_z = eBv_y = eaBh_{GW}\omega$, pointing along the wires (z-direction) thereby causing a current flow in the loop. For the frictionless motion of Cooper pairs in a superconducting wire with cross-section $S$, the velocity change at each quarter cycle of oscillation is $v_z = F_z/(m_e \omega)$, so that the current is $J = ev_z nS$. In turn, this current interacts with the magnetic field and generates an opposing (feedback) torque $\tau_{f.b.} \approx (ev_z Ba)aSn = JBa^2$. Here, only the major contribution from the parts of the frame along the z-direction was taken into account, so that $\tau_{f.b.} \approx eB^2 a^3 nAh_{GW}/m_e$. Obviously, $\tau_{f.b.} \leq \tau_z^{GW}$, and that means that for given parameters of the device, such as $M, a, r$, the value of the field $B$ cannot be set arbitrary high, or otherwise the frame (and the attached torus) will hardly move, and negligible current will be generated. Short of writing and solving the differential equation which treats the problem exactly, one can still proceed by optimizing the design. To that end the parameters should be chosen so that the magnitude of the reaction torque is half the original torque due to GW ("impedance matching"). Imposing this yields:

$$M = 2M_0 \equiv 2\frac{n\lambda_L^2 e^2 a^3 B^2 / m_e}{r^2 \omega^2 / 4} N_0. \tag{13}$$

In the last equation the number of turns in the coil, $N_0$, was taken into account. Also the conductor cross-section was set to $S = \lambda_L^2$, where $\lambda_L = (m_e/\mu_0 ne^2)^{1/2}$ is the London penetration depth (see, *e.g.*, [22]) – this choice justifies the assumed uniformity of the $B$-field and the resulting current density through the wire cross-section. Solving Eq. (13) for given $M, N_0, r, a$ and the GW frequency $\omega$ yields the optimal $B$:

$$B = \frac{r\omega\mu_0^{1/2}}{(2a)^{3/2}}\left(\frac{M}{N_0}\right)^{1/2}. \tag{14}$$

The condition (14) is important for determining strain sensitivity of the detector, so it useful to re-derive it in a different way. The magnetic energy generated by the current loop is given by:

$$E_{magn} = L_{ind}J^2/2 = (\delta\Phi)^2/(2L_{ind}) = (BA)^2 h_{GW}^2/(32L_{ind}), \tag{15}$$

where $L_{ind}$ is the self-inductance of the loop, and $J$ is the current induced in it via $\delta\Phi = BA\delta\alpha$. The kinetic energy imparted by the GW to the torus is:

$$E_{kin} = I\Omega^2/2 \approx M(\omega h_{GW} r)^2/32. \tag{16}$$

The criterion $E_{kin} \geq E_{magn}$ implies

$$M\omega^2 r^2 \geq (BA)^2/L_{ind}. \tag{17}$$

For an equilateral rectangular coil, as above, $A = a^2$. One can approximate rectangular inductance by circular: $L_{ind} \cong \mu_0 a [\ln(4a/\lambda_L) - 2] N_0$, thus obtaining:

$$M \geq \frac{B^2 a^3}{\mu_0 [\ln(4a/\lambda_L) - 2] N_0 r^2 \omega^2}. \qquad (18)$$

Equation (18), up to a dimensionless numerical-logarithmic factor, coincides with Eq.(14). Substitution of Eq. (14) into (9) yields

$$h_{GW}^{\min} \sim \frac{10^{-5} \phi_0}{(N_0 M a \mu_0)^{1/2} r \omega} \qquad (19)$$

for the optimized detector design. Some typical sensitivity curves based on Eq. (19) are shown in Fig. 4. Equation (19) implies that almost over the whole range of parameters considered in Fig. 4 the $B$-field does not exceed 1 Tesla. Unless the $B$-field is changed with frequency range selection, the response will not be optimized, and the straight lines in Fig. 4 will further bend upwards at the lower frequencies in the manner of the LIGO curve. Also the size of the detector determines the upper range of the frequency response. The time required for the mechanical rotation to reach the elements of the device is the size divided by the speed of sound in the material. This time should be shorter than the period of the GW. For the sizes considered, $a \sim r \sim 1 meter$, this implies an upper bound on the range of detectable GW frequencies of 1 to 5 kHz. This will induce an upward bending of the lines in Fig. 4 at higher frequencies (also similar to the LIGO curve).

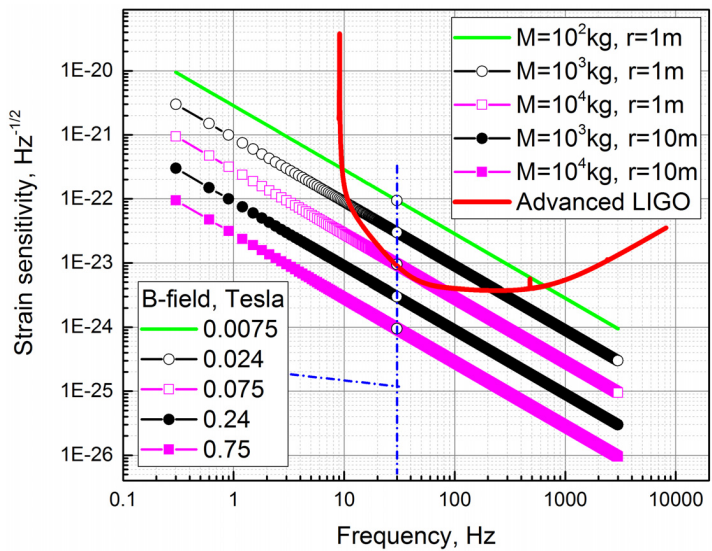

FIG. 4. Frequency dependence of strain sensitivity for a frame size $a=1$ *meter* vs. different values of torus radius $r$ and mass M. $N_0=10^4$. The optimal magnetic field B varies along each of the lines and grows with frequency and with mass. In particular, the B-values for a frequency of 30Hz are explicitly shown. The LIGO sensitivity (ZERO DET, high P [23]) is also shown.

## VII. Calibration

In complex multicomponent GW detectors, calibration is crucial in order to verify that the observed signal is indeed due to the GW. For example, LIGO can bounce pulses of light off the mirrors moving them by known calculable amounts. This amounts to perturbing in a controlled manner the path length in the interferometer so as to verify the correct final output.

We will next describe how a calibration could be achieved for our device. Calibration could be done for each part separately and for the whole system. For the latter calibration, a quadrupolar near-field gravitational perturbation can be generated by moving masses near the device. This is feasible thanks to the relatively small size of the proposed apparatus. While doing it, special precaution should be taken to exclude any concomitant non-gravitational forces. For example, the mechanical motion of masses may cause flow of air which will slightly deform the walls of the cryostat and create spurious influence. This acoustical coupling should be excluded by creating additional acoustic barrier via vacuum shells.

An important issue is the stability of the magnetic field. A superconducting solenoid with a closed loop coil structure with extremely stable supercurrents would yield no noticeable drift of the resulting B-field during the operation time. Special magnetic field sensors (*e.g.*, dedicated SQUIDs) may continually monitor the B-field and provide the required feedback in case of any drifts.

The calibration of the frame motion is possible by externally generating oscillatory motion of the coil. Care should be taken here not to exceed the region of linear response of the device and not to over flood the electronics.

The ability to look for coincidences with future signals from LIGO detectors allows excellent overall calibration of the suggested devices.

### VIII. Conclusion

To conclude, very efficient all-solid-state detectors may be achieved by converting the tidal acceleration of GW into rotational oscillatory motion of massive frames with subsequent transformation of this oscillatory motion into electrical signals. Typical meter-size detectors can be readily cooled to the required Kelvin or even sub-Kelvin temperature. Obvious advantages of the suggested modest size design are the possibility of orienting the instruments to maximize the signal from a selected source, and having duplicate detectors at various locations on Earth and in space. Mechanically the detector is non-resonant, however, electronically it can be tuned to very narrow bandwidths (via standard band-pass filtering techniques) to suppress the noise for the case of GW sources with well-defined frequency. Overall, the discussed concepts promise achieving a useful design and eventually tools for astrophysical exploration and studies of novel physical effects through the new observational window just opened by LIGO.

We are grateful to many our colleagues tor very useful discussions.